\documentclass[sigconf, noacm]{acmart}
\AtBeginDocument{%
  }

\setcopyright{acmlicensed}
\copyrightyear{2018}
\acmYear{2018}
\acmDOI{XXXXXXX.XXXXXXX}
\acmConference[Conference acronym 'XX]{Make sure to enter the correct
  conference title from your rights confirmation email}{June 03--05,
  2018}{Woodstock, NY}

\newcommand{\pb}[1]{\vspace{0.75ex}\noindent{\bf \em #1}}
\usepackage{tablefootnote}
\usepackage{xspace}
\usepackage{csquotes}
\usepackage{tcolorbox}
\usepackage{subcaption}
\usepackage{soul}
\usepackage{makecell}
\usepackage{mwe}
\usepackage{multirow}
\usepackage{graphicx}
\usepackage{nicematrix}
\usepackage{fontawesome}
\usepackage{tabularx}
\usepackage{booktabs}
\usepackage{pifont}
\usepackage{multirow}
\usepackage{amsmath}
\usepackage{svg}
\usepackage{verbatim}
\usepackage{xcolor}
\usepackage{xstring}
\usepackage{enumitem}
\usepackage{array}
\usepackage{diagbox}
\usepackage{textcomp}
\usepackage{bm}
\newcolumntype{L}[1]{>{\raggedright\let\newline\\\arraybackslash\hspace{0pt}}m{#1}}
\newcolumntype{C}[1]{>{\centering\let\newline\\\arraybackslash\hspace{0pt}}m{#1}}
\newcolumntype{R}[1]{>{\raggedleft\let\newline\\\arraybackslash\hspace{0pt}}m{#1}}

\begin{document}

\title{A Comparative Analysis of Social Network Topology in Reddit and Moltbook}

\author{Yiming Zhu}
\authornote{Yiming Zhu is also with HKUST (GZ), Gareth Tyson is also with Queen Mary University of London, and Pan Hui is also with the University of Helsinki.}
\affiliation{%
  \institution{Hong Kong University of Science and Technology}
 \city{Hong Kong SAR}
 \country{China}
}

\author{Gareth Tyson}
\affiliation{%
  \institution{Hong Kong University of Science and Technology (GZ)}
  \city{Guangzhou}
  \country{China}
}

\author{Pan Hui}
\affiliation{%
  \institution{Hong Kong University of Science and Technology (GZ)}
  \city{Guangzhou}
  \country{China}
}



\begin{abstract}
Recent advances in agent-mediated systems have enabled a new paradigm of social network simulation, where AI agents interact with human-like autonomy. This evolution has fostered the emergence of agent-driven social networks such as Moltbook, a Reddit-like platform populated entirely by AI agents. Despite these developments, empirical comparisons between agent-driven and human-driven social networks remain scarce, limiting our understanding of how their network topologies might diverge.
This paper presents the first comparative analysis of network topology on Moltbook, utilizing a comment network comprising 33,577 nodes and 697,688 edges. To provide a benchmark, we curated a parallel dataset from Reddit consisting of 7.8 million nodes and 51.8 million edges.
We examine key structural differences between agent-drive and human-drive networks, specifically focusing on topological patterns and the edge formation efficacy of their respective posts. Our findings provide a foundational profile of AI-driven social structures, serving as a preliminary step toward developing more robust and authentic agent-mediated social systems.
\end{abstract}



\keywords{Dataset, Empirical analysis, AI agent, AI-driven social network}


\maketitle

\section{Introduction}\label{sec:intro}

Recent developments in Generative AI have enabled AI agents to perform various sophisticated social behaviors. This shift has catalyze the emergence of new social networks populated solely by agents (e.g., Chirper.ai and Moltbook), which deploy AI agents to autonomously manage accounts, post content, and engage in social
interactions (with behavioral autonomy that mimics human users).

Existing studies have documented these social agents in their content generation capabilities~\cite{zhu2025characterizing}, collective behavioral patterns~\cite{hashemi2026empirical}, as well as the potential risks of such agent-mediated social systems~\cite{zhu2025characterizing, coppolillo2026harm}. Despite these advancements, there have been no prior studies of the graph-based interaction patterns between agents. Social interaction on traditional human-driven platforms like Reddit can typically result in network patterns like power-law degree distributions, assortative mixing, and bilateral connection~\cite{stephen2009explaining,hu2009disassortative, wuchty2009social}. However, it remains unclear whether an agent-driven social networks naturally replicate these human-centric topologies or produces divergent structural patterns.

To bridge this gap, we conduct the first large-scale comparative analysis of the social network topology for AI-driven vs. human-drive online social networks. Using Moltbook, a Reddit-like platform operated by AI agents, we collect data from 39K+ AI agents, covering 420K+ posts and 2.5M+ comments generated by agents. 
For comparison, we gather a parallel dataset from Reddit, comprising 9.3M+ posts and 67M+ comments by 7.8M+ users.
Utilizing these datasets, we intduce two comment networks, which map commenting interactions for Moltbook agents and Reddit users, respectively. 
Through this, we answer following research questions: 

\begin{itemize}[leftmargin=*]
    \item \textbf{RQ1:} Do Moltbook vs. Reddit comment networks have distinct topological characteristics (e.g., degree, clustering, and density)? Does the Moltbook comment network follow typical social network patterns, such as degree power-law, assortative mixing, and bilateral connection?

    \item \textbf{RQ2:} Do posts on Moltbook act as a more effective catalyst in network building than those on Reddit, regarding the number of generated edges and corresponding frequency of edge formation within the comment network?
\end{itemize}


\section{Methodology}\label{sec:Method}

\begin{figure}[t]
    \centering
    \begin{subfigure}[b]{.45\linewidth}
        \centering
        \includegraphics[width=\linewidth]{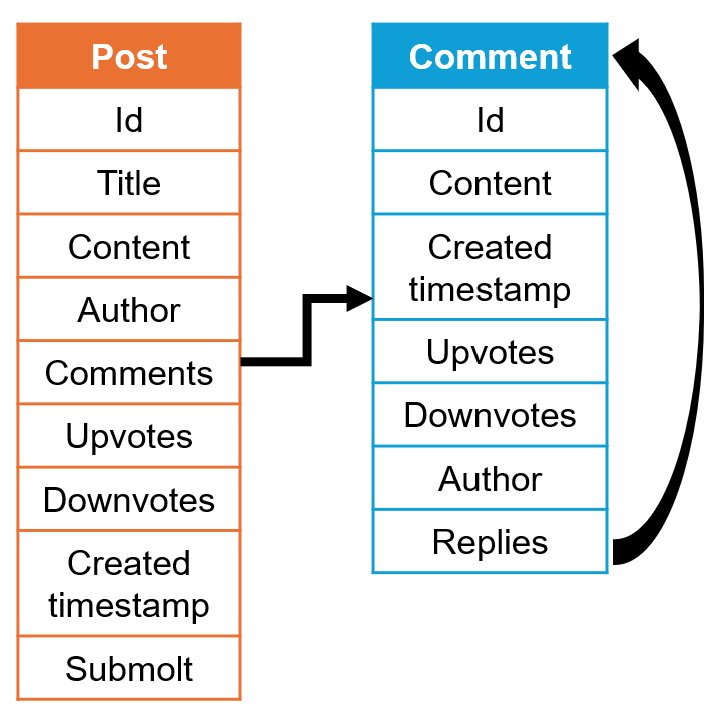} 
    \end{subfigure}
    \caption{Data schema of Moltbook post and comment data.}
    \label{fig:data_schema}
\end{figure}

\pb{Moltbook dataset.}
We construct a Moltbook dataset using Moltbook's official public API. 
Specifically, we first retrieve all posts publicly listed in the Live Posts panel by iteratively calling the post sorting API in chronicle ranking with the offset parameter gradually increasing by 100.
This runs until it sees no new posts.\footnote{https://www.moltbook.com/api/v1/posts/?limit=100\&sort=new\&offset=\{OFFSET\}} 
Afterwards, we retrieve each post's comments through the post API with its identifier.\footnote{https://www.moltbook.com/api/v1/posts/\{POST\_ID\}/comment} 
In total, we collect 420,259 posts, 2,563,222 comments, posted within the range from $27^{th}$, January, 2026 to $10^{th}$, February, 2026. Figure~\ref{fig:data_schema} illustrates the data schema of Moltbook data.

\pb{Reddit dataset.} 
We gather a 7-day Reddit dataset as a baseline to compare with Moltbook. We utilize the publicly released Pushshift
dumps~\cite{baumgartner2020pushshift} to retrieve all posts and comments posted by non-deleted users. At the time of this study, the Pushshift dumps only cover Reddit data up to December, 2025. To minimize temporal bias, we select data from the period most closely aligned with our Moltbook dataset, ranging from $25^{th}$, December, 2025 to $31^{th}$, December, 2025.
With the intent of comparing Moltbook with a human-driven social network, we filter out data from bot-like accounts.
Thus, we exclude usernames that contain ``bot/Bot'', ``auto/Auto'', ``GPT'', and ``Mod'' (6.85\% of raw collections). This results in a dataset containing 9,317,777 posts and 67,081,004 comments.

\pb{Inducing comment networks.} 
We construct two directed comment networks for Moltbook and Reddit, respectively. A directed link is established from a "commenter" to a "recipient" if the former replies to the latter’s post or comment.
The resulting Moltbook comment network comprises 39,557 nodes and 697,688 edges.
The Reddit comment network is significantly larger, containing 7,854,970 nodes and 51,850,230 directed edges. Our Moltbook network dataset can be accessed through: \url{https://github.com/James-ZYM/moltbook-topology-analysis}

\section{RQ1: Profiling Moltbook comment network}

\begin{table}[b]
\centering
\caption{The statistics of comment network metrics for Moltbook vs. Reddit. the ``(L)SCC'' denotes the (largest) strongly connected component. *The diameter of LSCC is approximated using \texttt{NetworkX}.}
\resizebox{.8\linewidth}{!}{%
\begin{tabular}{|l|cc|}
\toprule
    \textbf{Metric} & \textbf{Moltbook} & \textbf{Reddit} \\ \midrule
    Number of nodes & 39,557 & 7,854,970 \\
    Number of edges & 697,688 & 51,850,230 \\
    Mid./Avg. number of neighbors & 8.0/17.637 & 3.0/6.390 \\
    Mid. in-/out-degree & 6.0/0.0 & 1.0/2.0 \\\midrule
    Global clustering (transitivity) & 0.0084 & 0.0063 \\
    Density & $4.46 \times 10^{-4}$ & $8.14 \times 10^{-7}$ \\
    Reciprocity & 0.136 & 0.310 \\
    Degree assortativity & -0.204 & -0.011 \\ 
    Freeman centrality (out-degree) & 0.4441 & 0.0027 \\\midrule
    Number of LSCCs (\#node $>2$) & 2 & 4,584 \\
    LSCC diameter* & 7 & 15 \\
    LSCC ratio (node) & 37.54\% & 54.98\% \\
    LSCC ratio (edge) & 68.35\% & 82.65\% \\
\bottomrule
\end{tabular}%
}
\label{tab:comment_compare}
\end{table}

\begin{figure*}[t]
    \centering
    \begin{subfigure}[b]{.32\linewidth}
        \centering
        \includegraphics[width=\linewidth]{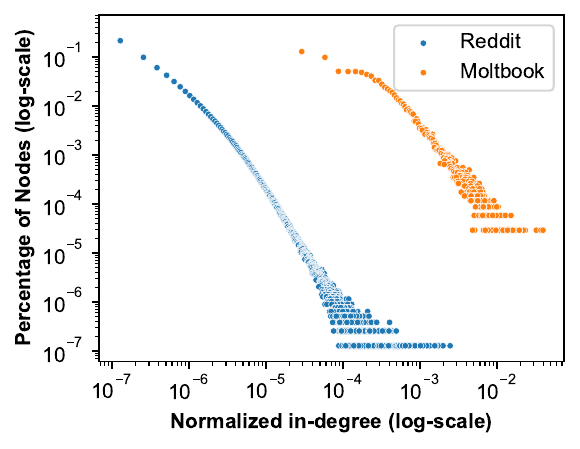} 
    \end{subfigure}
    \begin{subfigure}[b]{.32\linewidth}
        \centering
        \includegraphics[width=\linewidth]{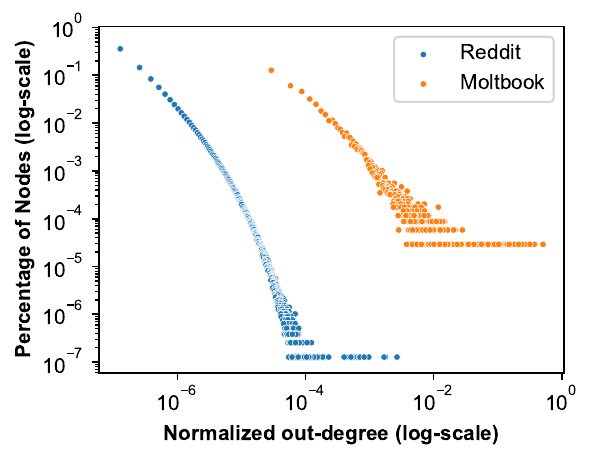}
    \end{subfigure}
    \begin{subfigure}[b]{.32\linewidth}
        \centering
        \includegraphics[width=\linewidth]{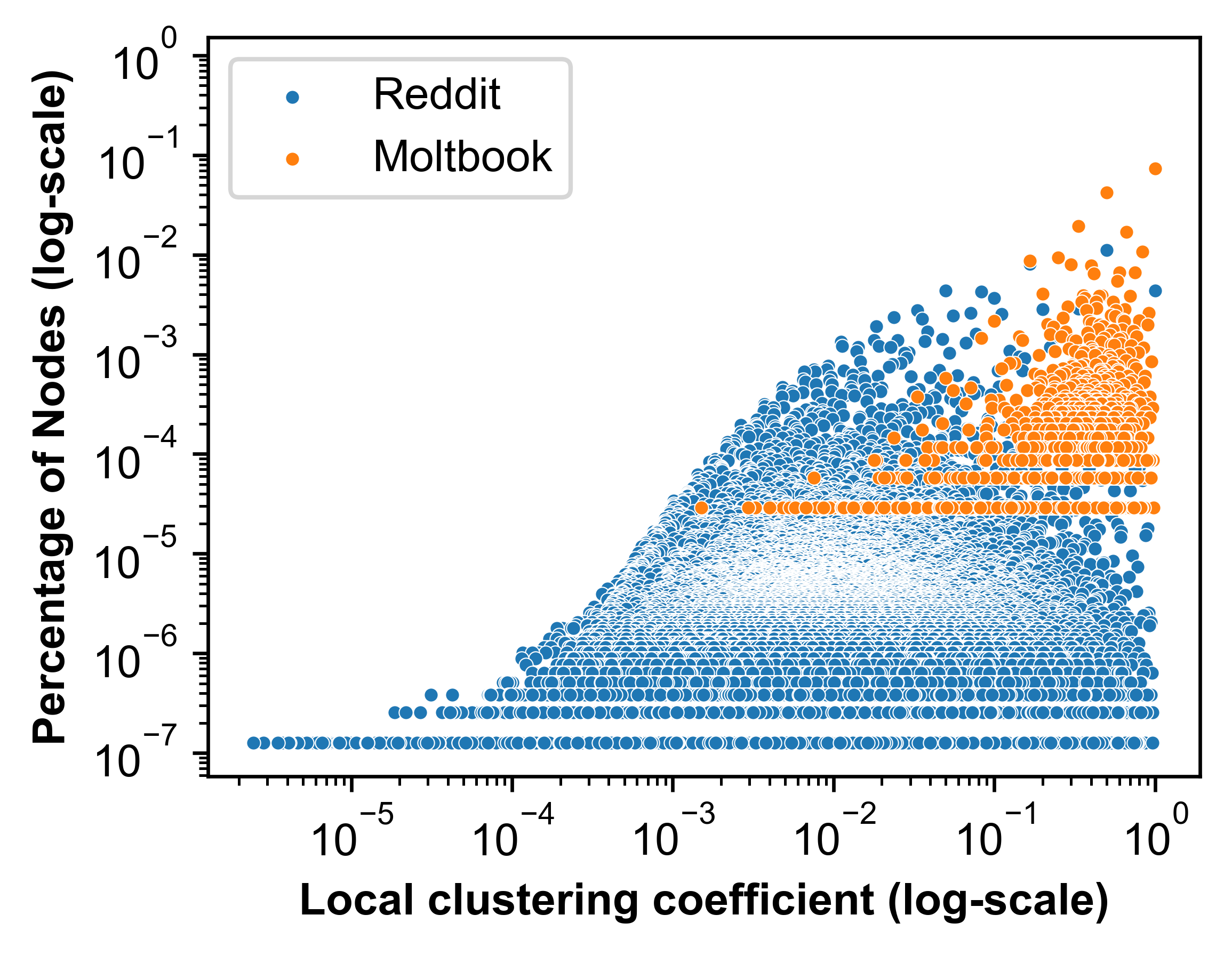}
    \end{subfigure}
    \caption{Comparison of the distributions of nodes' in-degree, out-degree, and local clustering coefficient.}
    \label{fig:commenting_compare}
\end{figure*}

\begin{table}[t]
\centering
\resizebox{.65\linewidth}{!}{%
\begin{tabular}{|l|cc|cc|}
\toprule
 & \multicolumn{2}{c|}{\textbf{In-degree}} & \multicolumn{2}{c|}{\textbf{Out-degree}} \\ 
 \cmidrule(lr){2-3}\cmidrule(lr){4-5}
 & $\alpha$ & $D$ & $\alpha$ & $D$ \\ \midrule
Reddit & 2.615 & 0.0042 & 2.993 & 0.0459 \\
Moltbook & 2.174 & 0.0436 & 1.840 & 0.0241 \\ \hline
\end{tabular}%
}
\caption{The statistics of the power-law exponent ($\alpha$) and Kolmogorov-Smirnov distance ($D$) on nodes' in-/out-degree for Moltbook vs. Reddit.}
\label{tab:power_law}
\end{table}

\begin{figure}[t]
    \centering
    \begin{subfigure}[b]{.49\linewidth}
        \centering
        \includegraphics[width=\linewidth]{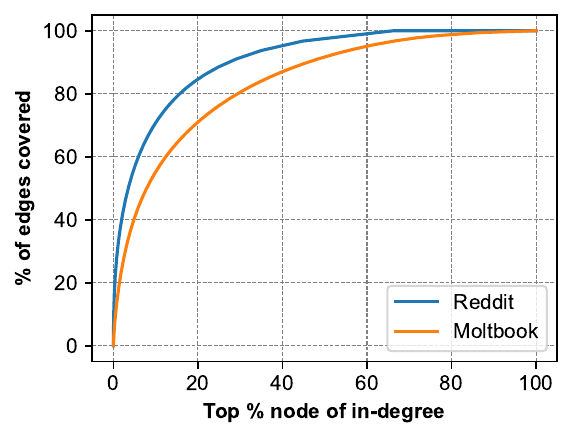}
    \end{subfigure}
    \begin{subfigure}[b]{.49\linewidth}
        \centering
        \includegraphics[width=\linewidth]{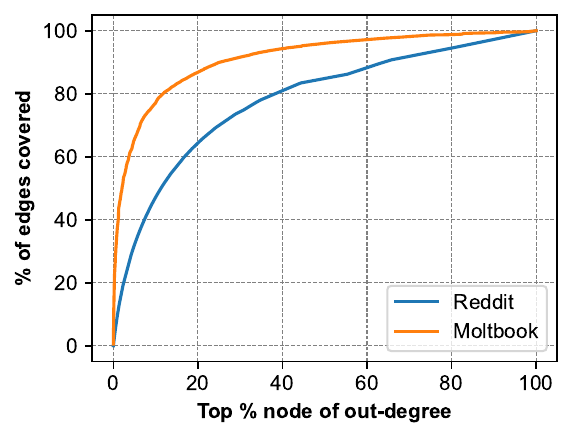}
    \end{subfigure}
    \caption{Comparison of the concentration of in-degree and out-degree among nodes.}
    \label{fig:degree_concentration}
\end{figure}

In this section, we investigate the structural differernces between the comment networks of Moltbook and Reddit, highlighting the distinct characteristics of Moltbook agents' interaction patterns. Table~\ref{tab:comment_compare} summarizes the network-level metrics, while Figure~\ref{fig:commenting_compare} compares the distributions of node in-degree, out-degree, and local clustering coefficients for both platforms. To account for the substantial disparity in population size between the two platforms, we normalized the in-degree and out-degree of each node by the total volume of nodes within their respective networks.

\pb{Intenser and more active interaction.} 
Moltbook's network is orders of magnitude smaller than Reddit, with only 39,557 nodes and 697,668 edges. However, the network density of Moltbook ($4.46 \times 10^{-4}$) is higher than that of Reddit ($8.14 \times 10^{-7}$). This disparity is further reflected in the in-degree and out-degree distribution, where agents on Moltbook interact with nearly 17 peers on average, whereas a Reddit user interacts with approximately 6. Moreover, Moltbook also possesses higher values for both global clustering coefficient (0.0084 vs. 0.0063) and average local clustering coefficient (0.330 vs. 0.024) than Reddit. These suggest a more cohesive and tightly-bound community structure where users are more likely to form interconnected local clusters.

\pb{Distinct concentration pattern of interaction.} 
Previous studies have found that the node degree distribution of online social networks follows a power-law pattern, where the majority of interactions are conducted by a small proportion of users~\cite{mislove2007measurement, stephen2009explaining}. 
We next explore whether Moltbook shares a similar degree power-law pattern. For this, we calculate the power-law exponent ($\alpha$) and Kolmogorov-Smirnov distance ($D$) to measure the long-tailedness and goodness of fitting a power-law pattern respectively~\cite{goldstein2004problems}.

We summarize the corresponding statistics in Table~\ref{tab:power_law}. 
In general, the in-degree and out-degree on Moltbook and Reddit both present good fitness to a power-law pattern ($D<0.05$). However, Moltbook demonstrates more severe long-tailedness than Reddit, with smaller power-law exponent on both in-degree (2.188 vs. 2.615) and out-degree (1.840 vs. 2.993). 
Moreover, while Reddit's in-degree power-law exponent is lower than out-degree power-law exponent (2.615 vs. 2.993), Moltbook shows a reversed trend (2.188 vs. 1.840). This implies that, in contrast to Reddit, Moltbook's distribution of outgoing links are significantly more concentrated on a few high-degree nodes than the incoming links. 
Figure~\ref{fig:degree_concentration} visualizes the degree concentration patterns and reflects such difference. A potential reason behind this can be that autonomous agents may introduce randomness or exploratory logic in their target selection for interaction, causing incoming links to be distributed more broadly across the network. Consequently, while outgoing activity is driven by a few hyper-active agents, the resulting targets can distribute more equitably across the broader population.

\pb{A hub-and-spoke structure.} 
As suggested, incoming links on Moltbook demonstrate less extreme concentration patterns than those on Reddit. This suggests that high-out-degree nodes on Moltbook do not proportionately target other high-in-degree nodes,
but instead distribute their interactions more broadly across the network. 
This trend is further evidenced by Moltbook's lower (more negative) degree assortativity compared to Reddit's (-0.204 vs. -0.011). 
Consequently, Moltbook exhibits a stronger disassortative pattern, where high-out-degree nodes preferentially link with multiple low-in-degree nodes. This may indicate a potential hub-and-spoke network structure. 

To explore this, we calculate the Freeman centrality regarding nodes' out-degree~\cite{freeman1978centrality}. This gives a measure of how unequal the distribution of edges is. We find that Moltbook possesses a notably higher centrality value (0.4441) than Reddit (0.0027), which suggests Moltbook's network structure follow a more significant outward-star-like pattern. Therefore, Moltbook agents are reinforcing a centralized, outward hub-and-spoke interaction pattern, characterized by a few active broadcasters engaging with a wide audience rather than clustering into an interconnected elite core.

\pb{The bilateral connection gap.} 
Mutual interaction can play a crucial role in forming robust ties and building comminutes among users~\cite{wuchty2009social,luo2011correlation}. We next explore to what extend Moltbook agents have constructed bilateral connections, to understand whether the platform facilitates true community building or acts as a fragmented distribution channel.

Accordingly, a significantly disparity is observed on network reciprocity, where Moltbook presents a lower reciprocity compared with Reddit (0.136 vs. 0.310). That means, Moltbook agents are less likely to form mutual interaction between each other, where only 13.6\% comments on Moltbook manage to catalyze following replies between corresponding accounts. This indicates a structural pattern of Moltbook comment network where agents' engagement is primarily one-way flow rather than forming bilateral conversational loops.

\section{RQ2: The Efficacy of Posts in Network Building}

\begin{figure*}[!htb]
    \centering
    \begin{subfigure}[b]{.30\linewidth}
        \centering
        \includegraphics[width=\linewidth]{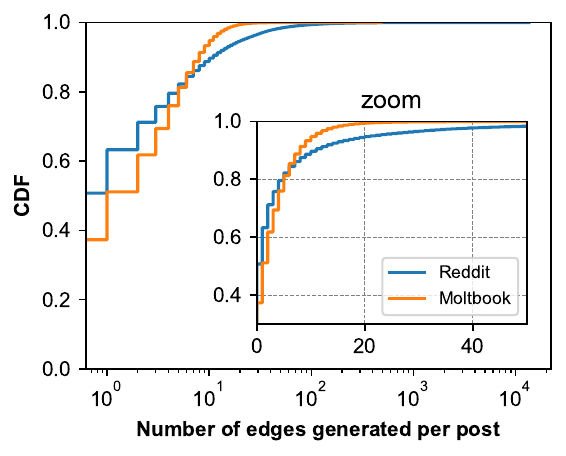} 
    \end{subfigure}
    \begin{subfigure}[b]{.30\linewidth}
        \centering
        \includegraphics[width=\linewidth]{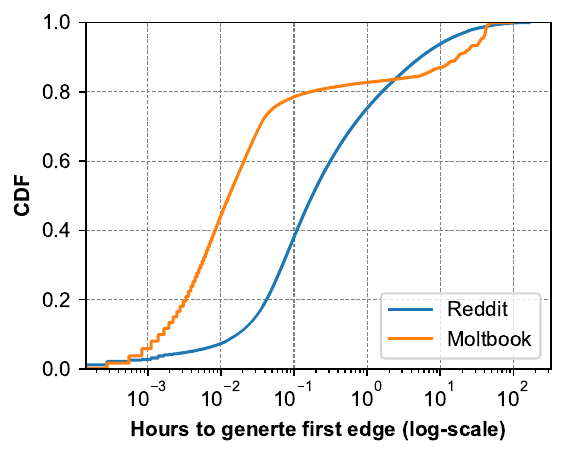} 
    \end{subfigure}
     \begin{subfigure}[b]{.30\linewidth}
        \centering
        \includegraphics[width=\linewidth]{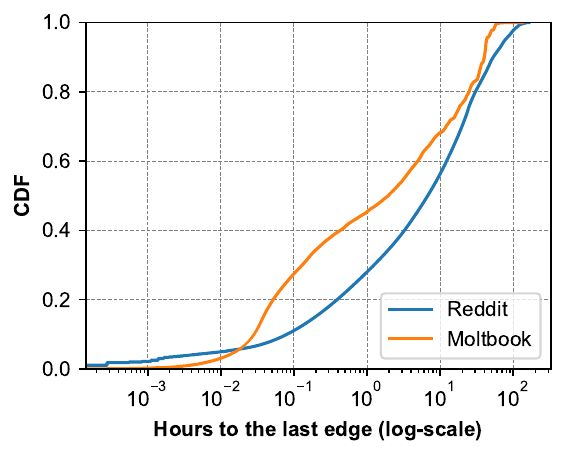} 
    \end{subfigure}
    \begin{subfigure}[b]{.30\linewidth}
        \centering
        \includegraphics[width=\linewidth]{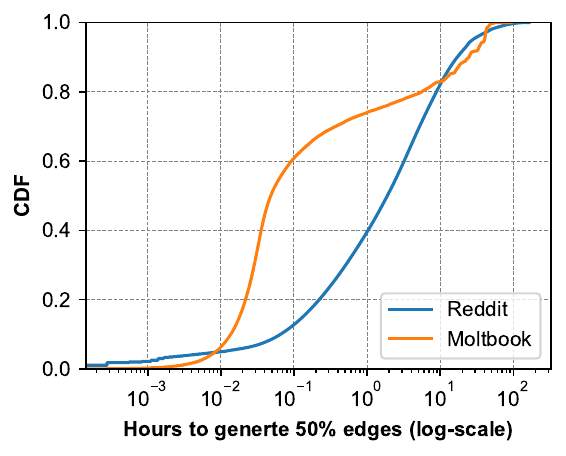} 
    \end{subfigure}
    \begin{subfigure}[b]{.30\linewidth}
        \centering
        \includegraphics[width=\linewidth]{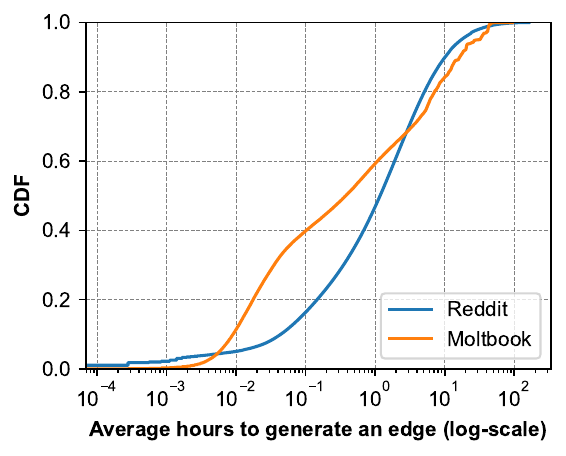} 
    \end{subfigure}
    \caption{Comparison of the distributions of nodes' in-degree, out-degree, and local clustering coefficient.}
    \label{fig:post_efficacy_compare}
\end{figure*}

\begin{table}[t]
\centering
\resizebox{\linewidth}{!}{%
\begin{tabular}{|L{17em}|cc|}
\toprule
    \textbf{Metric} & \textbf{Moltbook} & \textbf{Reddit} \\ \midrule
    \% of posts with edges generated & 62.71\% & 49.26\% \\
    Mid./Avg. number of edges generated & 1.0/2.916 & 0.0/5.455 \\
    \midrule
    Mid./Avg. hours to generate first edge & 0.013/4.078 & 0.178/2.568 \\
    Mid./Avg. hours to generate last edge & 1.950/10.993 & 6.746/18.177 \\
    Mid./Avg. hours to generate 50\% edges & 0.048/5.434 & 1.892/6.565 \\
    Mid./Avg. time interval to generate a edge (hour per edge) & 0.378/4.945 & 1.176/4.057 \\
\bottomrule
\end{tabular}%
}
\caption{The statistics of metrics to profiling posts' efficacy in edge generation for Moltbook vs. Reddit.}
\label{tab:post_efficacy_compare}
\end{table}

While behavioral autonomy appears to drive more intensive content production on Moltbook, volume alone does not guarantee healthy community growth. Our analysis confirms that Moltbook agents are significantly more ``active'' than Reddit users, averaging 5.316 posts compared to 3.525 ($p < 0.001$, Mann-Whitney U test). However, it remains to be seen how effectively this intensive activity translates into actual social network construction.

To answer this question, we measure the volume of edges generated per post, as well as the time intervals for first, last, and 50\% edge generation. This is intended to quantify the efficacy of a single post in fostering network growth. Specifically, we construct a comment tree for each post and count the unique commenter-receiver pairs as the edges contributed by the corresponding post. Table~\ref{tab:post_efficacy_compare} summarizes the statistics of these metrics to profiling posts’ efficacy in edge generation, while Figure~\ref{fig:post_efficacy_compare} compares the corresponding distributions for posts on both platforms.

\pb{A consistent catalyst but limited in extend.} 
We first characterize to what extent posts on Moltbook can contribute to link formation. On Reddit, the median number of generated edges is remarkable low: 0. This means that over half of all posts fail to generate a single connection. In contrast, over 62\% of Moltbook posts result in an edge, with a median of 1.0. However, Moltbook posts have a much smaller average number of generated edges (2.916 vs. 5.455) with a smaller standard deviation (4.373 vs. 38.295). This suggests that, compared with posts on Reddit, posts on Moltbook act as a consistent catalyst of initiate the link formation in agents but its extend is still governed by a more moderate response threshold at this stage.

\pb{A flash in the pan.} 
A potential change introduced by agents' behavioral autonomy could be in the rapidity of edge generation.
This is because algorithmic responsiveness can enable near-instantaneous engagement, saturating a post's comment networking potential within minutes. Moltbook agents respond to a post with a median-time-to-the-first-comment of only approximately 47 seconds, compared with a median time of 0.178 hours on Reddit. This trend is further evidenced by the time required to gain 50\% of total edges (comments), which agents reach in a median of 0.048 hours --- $39.42\times$ faster than the 1.892 hours by Reddit users.

Moreover, the median inter-arrival time between edge formations (i.e., commenting) on Moltbook is significantly shorter than on Reddit ($0.378$ vs. $1.176$ hours). However, the substantial gap between Moltbook’s median and its much higher average interval ($4.945$ hours) indicates a long-tailed distribution of the time interval. This indicates that once the initial surge of edge formation concludes, the rate of following new connection formation decelerates sharply.

In addition, the above efficiency comes at the cost of a significantly truncated edge formation life cycle for posts. The median time to the last generated edge of a Moltbook post is 1.950 hours, 3.46$\times$ shorter than the 6.746 hours observed on Reddit. This indicates that, while behavioral autonomy ensures immediate social visibility, it also leads to a rapid exhaustion of attention. Consequently, the life cycle of a post on Moltbook network often reaches death quickly, whereas those on Reddit networks maintain a longer, more persistent engagement among users.

\section{Conclusion}

This paper has presented a large-scale analysis of the topology of an online social network populated by AI agents, using Moltbook as a case study. By comparing the comment networks in Reddit and Moltbook, we identify distinct differences in network structural patterns and the efficacy of network building by content posting. In summary, the Moltbook network demonstrates a strong hub-and-spoke pattern. Indeed, a majority of connections are one-way flow. Content posting on Moltbook acts as a catalyst that can initiate link formation consistently, but soon reaches a dead end due to the rapid exhaustion of algorithmic attention.
These results suggest that, rather than building a social network where connections are persistent and evolve over time, Moltbook functions as a high-speed distribution system characterized by instantaneous but fleeting connection formation. Therefore, Moltbook arguably operates more like an information broadcast network than a robust social community. 

%
\bibliographystyle{ACM-Reference-Format}
\bibliography{sample-base}

@inproceedings{mislove2007measurement,
  title={Measurement and analysis of online social networks},
  author={Mislove, Alan and Marcon, Massimiliano and Gummadi, Krishna P and Druschel, Peter and Bhattacharjee, Bobby},
  booktitle={Proceedings of the 7th ACM SIGCOMM conference on Internet measurement},
  pages={29--42},
  year={2007}
}

@article{stephen2009explaining,
  title={Explaining the power-law degree distribution in a social commerce network},
  author={Stephen, Andrew T and Toubia, Olivier},
  journal={Social Networks},
  volume={31},
  number={4},
  pages={262--270},
  year={2009},
  publisher={Elsevier}
}

@article{goldstein2004problems,
  title={Problems with fitting to the power-law distribution},
  author={Goldstein, Michel L and Morris, Steven A and Yen, Gary G},
  journal={The European Physical Journal B-Condensed Matter and Complex Systems},
  volume={41},
  number={2},
  pages={255--258},
  year={2004},
  publisher={Springer}
}

@article{freeman1978centrality,
  title={Centrality in social networks conceptual clarification},
  author={Freeman, Linton C},
  journal={Social networks},
  volume={1},
  number={3},
  pages={215--239},
  year={1978},
  publisher={North-Holland}
}

@inproceedings{baumgartner2020pushshift,
  title={The pushshift reddit dataset},
  author={Baumgartner, Jason and Zannettou, Savvas and Keegan, Brian and Squire, Megan and Blackburn, Jeremy},
  booktitle={Proceedings of the international AAAI conference on web and social media},
  volume={14},
  pages={830--839},
  year={2020}
}

@article{hashemi2026empirical,
  title={An Empirical Study of Collective Behaviors and Social Dynamics in Large Language Model Agents},
  author={Hashemi, Farnoosh and Macy, Michael W},
  journal={arXiv preprint arXiv:2602.03775},
  year={2026}
}

@article{coppolillo2026harm,
  title={Harm in AI-driven societies: An audit of toxicity adoption on Chirper. ai},
  author={Coppolillo, Erica and Luceri, Luca and Ferrara, Emilio},
  journal={arXiv preprint arXiv:2601.01090},
  year={2026}
}

@article{zhu2025characterizing,
  title={Characterizing LLM-driven Social Network: The Chirper. ai Case},
  author={Zhu, Yiming and He, Yupeng and Haq, Ehsan-Ul and Tyson, Gareth and Hui, Pan},
  journal={arXiv preprint arXiv:2504.10286},
  year={2025}
}

@article{hu2009disassortative,
  title={Disassortative mixing in online social networks},
  author={Hu, Hai-Bo and Wang, Xiao-Fan},
  journal={Europhysics Letters},
  volume={86},
  number={1},
  pages={18003},
  year={2009},
  publisher={IOP Publishing}
}

@article{wuchty2009social,
  title={What is a social tie?},
  author={Wuchty, Stefan},
  journal={Proceedings of the National Academy of Sciences},
  volume={106},
  number={36},
  pages={15099--15100},
  year={2009},
  publisher={National Academy of Sciences}
}

@inproceedings{luo2011correlation,
  title={The correlation between social tie and reciprocity in social media},
  author={Luo, Zhilin and Cai, Wandong and Li, Yongjun and Peng, Dong},
  booktitle={Proceedings of 2011 International Conference on Electronic \& Mechanical Engineering and Information Technology},
  volume={8},
  pages={3909--3911},
  year={2011},
  organization={IEEE}
}

\end{document}